# Smart Sustainable Agriculture (SSA) Solution Underpinned by Internet of Things (IoT) and Artificial Intelligence (AI)

Eissa Alreshidi[1]

Assistant Professor at College of Computer Science & Engineering, University of Hail, Hail, P.O. Box 2440, Saudi Arabia

*Abstract*—The Internet of Things (IoT) and Artificial Intelligence (AI) have been employed in agriculture over a long period of time, alongside other advanced computing technologies. However, increased attention is currently being paid to the use of such smart technologies. Agriculture has provided an important source of food for human beings over many thousands of years, including the development of appropriate farming methods for different types of crops. The emergence of new advanced IoT technologies has the potential to monitor the agricultural environment to ensure high-quality products. However, there remains a lack of research and development in relation to Smart Sustainable Agriculture (SSA), accompanied by complex obstacles arising from the fragmentation of agricultural processes, i.e. the control and operation of IoT/AI machines; data sharing and management; interoperability; and large amounts of data analysis and storage. This study firstly, explores existing IoT/AI technologies adopted for SSA and secondly, identifies IoT/AI technical architecture capable of underpinning the development of SSA platforms. As well as contributing to the current body of knowledge, this research reviews research and development within SSA and provides an IoT/AI architecture to establish a Smart, Sustainable Agriculture platform as a solution.

## I. Introduction

Agriculture forms a critical activity vital to the survival of humanity for approximately many thousands of years [1]. This relationship has resulted in the advancement of agricultural activities, initially through the time-consuming methods of traditional agriculture [2]. The current recent rapid increase in in the global population (predicted to rise to 8.9 billion by 2050) has now led to an urgent need to balance demand and supply through the use of new technologies [3] to increase food production [4, 5]. This development places pressure on natural resources, with agriculture now consuming 70% of the world's fresh water supply for the purposes of irrigation. Limited resources and the impact of climate change will therefore lead to considerable challenges in producing sufficient high quality food to support the population [6]. Smart Agricultural is a global initiative to preserve resources and maintain sustainable agriculture [7]. Recently, researchers have adopted the Internet of Things (IoT) [8, 9], with a number of studies emphasizing the adoption and implementation of IoT in agriculture, farming, and irrigation [10]. Around the globe, many private companies and organizations are now focusing on investigating new technologies to create a smarter agriculture environment. These include mechanical and economic aspects, engineering, food retailers and computing. However, agricultural processes are fragmented, resulting in a number of issues, i.e. difficulties in operating and managing smart machines, data sharing and management, data analysis and storage [11, 12]. It is therefore important to facilitate cooperation when developing standards for smart agriculture, while also enhancing interoperability among different stakeholders, systems and technologies [13].

The use of IoT and AI technologies has the potential to result in a positive transformation of traditional agriculture [3], including: (a) improved use of data collected from smart agriculture sensors; (b) managing and governing the internal processes within the smart agriculture environment (including the management of the harvesting and storage of crops); (c) waste reduction and cost saving; (d) increasing business efficiency by means of automating traditional processes; and (e) improving the quality and volume of products [14]. A major challenge is to provide farmers with the required information in a rapid manner [15]. AI therefore has significant potential to address the urgent challenges faced by traditional agriculture. There has, over previous decades, been considerable research and application of AI, including in: (a) smart agriculture; (b) robotics; (c) agricultural optimization management; (d) automation; (e) agricultural expert systems; (f) agricultural knowledge-based systems; and (g) decision support systems [16].

There remains a lack of research and development in relation to Smart Sustainable Agriculture (SSA), accompanied by complex obstacles arising from the fragmentation of agricultural processes, i.e. the control and operation of IoT/AI machines; data sharing and management; interoperability; and large amounts of generated data analysis and storage. Therefore, this study firstly, explores existing IoT/AI technologies adopted for SSA and secondly, establishes an IoT/AI technical architecture to underpin SSA platforms, in order to tackle fragmentation in traditional agriculture processes and enrich the research and development of future smart agriculture worldwide via establishment of a Smart, Sustainable Agriculture platform as a solution.

There now follows an outline of the methodology underpinning this research, supported by related work highlighting the history of smart agriculture, smart and advanced computing technologies and examples of IoT/AI technologies in current agricultural practices. This is supported by an in-depth discussion of Smart Agriculture and IoT/AI SSA technical architecture, along with the most significant outcomes from this study. The paper finishes with concluding remarks and plans for future work.

## II. BACKGROUND

Due to the lack of literature concerning the development of IoT frameworks for SSA, this study focuses on: (a) the history of smart agriculture, its potential and challenges; (b) smart and advanced computing technologies; and (c) existing smart, sustainable agricultural frameworks.

### A. Smart Sustainable Agriculture

There has recently been considerable research into SSA, employing various different terms, including Precision Farming, Smart Irrigation and Smart Greenhouse. This paper commences with an examination of these concepts to determine the definition of SSA used in this study.

Precision Farming refers to a method of managing farms and conserving resources through the use of IoT and Information and Communication Technologies (ICT). It obtains real-time data concerning the condition of farm elements, (i.e. crops, soil and air) to protect the environment while ensuring profits and sustainability [15]. Smart Irrigation is a method of improving the efficiency of irrigation processes and reducing water losses, while conserving existing water resources using IoT-based smart irrigation systems [16]. Drones are employed in many agricultural applications, including monitoring field crops and livestock, and scanning large areas, while sensors on the ground collect a huge range of information [13]. Smart greenhouses promote the cultivation of crops with the least degree of human intervention possible, through use of continuously monitored climatic conditions (i.e. humidity, temperature, luminosity and soil moisture), triggering automated actions based on the evaluated changes and implementing corrective action to maintain the most beneficial conditions for growth [17].

Farm Management Systems (FMS) can assist farmers with a variety of collected information, by managing and controlling various tracking devices and sensors. The collected information is analysed for the undertaking of complex decision-making tasks before being placed in a storage medium. This enables the use of the most effective agricultural data analysis practices [18]. Soil Monitoring Systems help to track and improve the quality of soil through the monitoring of its physical, chemical, and biological properties. Livestock monitoring systems provide real-time assessment of the productivity, health and welfare of livestock, to promote the health of animals [19]. The IoT/AI SSA platform Cloud offers real-time information to farmers to facilitate decision-making and reduce operational costs, while at the same time enhancing productivity. Following a review of a considerable amount of research, we define SSA as the utilization process of IoT/AI technologies to establish, monitor, manage, process and analyse data generated from various agricultural resources, such as field, crops, livestock and others to ensure the sustainability and quality of agricultural products and further enrich decision-making taken by stakeholders.

### B. Smart and Advanced Computing Technologies

This section provides an overview of appropriate technologies underpinning the development of smart, sustainable agriculture platforms, including: IoT; Big Data Analytics (BDA); Cloud Computing (CC); Mobile Computing (MC); and Artificial Intelligence (AI):

*1) Internet of Things (IoT):* IoT is a technology aimed at connecting all intelligent objects within a single network, i.e. the Internet. It involves all kinds of computer technologies, both (a) hardware (i.e. intelligent boards and sensors) and (b) software (i.e. advanced operating systems and AI algorithms). Its primary target is the establishment of applications for devices, in order to enable the monitoring and control of a specific domain. It has been widely adopted in many areas, i.e. industrial business processes; home machines; health applications; and smart homes and cities. IoT connectivity encompasses people, machines, tools and locations, aiming to achieve different intelligent functions from data sharing and information exchange [17]. However, it is primarily used in agriculture for management of agricultural products within gathered real-time data, alongside: (1) searching; (2) tracking; (3) monitoring; (4) control; (5) managing; (6) evaluating; and (7) operations within a supply chain [1, 9].

*2) Big Data Analytics (BDA):* BDA refers to the large volume of data gathered from different datasets sources over a long period of time, i.e. sensor, Internet and business data. The datasets used in this technology surpass the computational and analytical capabilities of typical software applications and standard database infrastructure. Its primary task is to capture, store, analyze and search for data, as well as seeking to identify concealed patterns in the gathered data. Thus, BDA involved the utilization of: (a) tools, (i.e. classification and clustering); (b) techniques, (i.e. data mining, machine learning and statistical analysis); and (c) technologies (i.e. Hadoop and spark). These go beyond traditional data analytical approaches, being employed to extract beneficial knowledge from a considerable amount of data, in order to facilitate timely and accurate decision–making [17]. However, the use of BDA in agriculture focusses on management of the supply chain of agricultural products, in order to enhance decision-making and minimize the cost of production cost. It is also employed for the analysis of the properties of different types of soil for classification and further enhancement. Furthermore, it is useful for the improved prediction and production of crops.

*3) Cloud Computing (CC):* CC has is a recent and rapidly growing phenomenon within IT [18]. The Cloud is not restricted to a particular business domain, but has been implemented to underpin and support various software applications and platforms [19]. It offers easy access to the Cloud provider's high-performance and storage infrastructure over the Internet, with one of its main benefits being to conceal from users the complexity of IT infrastructure management [20] [21]. NIST [22] defined CC as "a model for enabling convenient, on-demand network access to a shared pool of configurable computing resources (e.g. networks, servers, storage, applications and services) that can be rapidly provisioned and released with minimal management effort or service provider interaction". The Cloud can be seen as high virtualization method for datacenter infrastructure distributed over a wide geographical area, linked by means of high bandwidth network cables providing a variety of virtualized services. These include entire infrastructures, as well as small software applications and different types of services, i.e. high-performance computing and large scalable storage services based on a pay-per-use model. CC can be divided into four main layers: (1) hardware; (2) infrastructure; (3) platform; and (4) application [23]. The delivery of Cloud services can generally be divided into three different models: (1) Infrastructure-as-a-Service (IaaS); (2) Platform-as-a-Service (PaaS); and (3) Software-as-a-Service (SaaS) [24, 25]. CC is considered the most effective method of storing agricultural data, along with IoT [1].

*4) Mobile Computing (MC):* MC refers to infrastructure in which data processing and data storage take place externally to the mobile device [26]. MC applications transfer computing power, processing and data storage from mobile devices in the Cloud [27, 28]. MC has had a considerable impact on modern daily life, due to the availability and low cost of purchasing and communication. It is now widely used in every field, including the agricultural sector [29], in which MC systems collect and send daily data to farmers, informing them of both the production status and weather conditions [29]. It is crucial to use automatic Radio Frequency Identification (RFID) efficient traceability systems to store and access data on electronic data chips in a more rapid and accurate manner. It has been primarily applied to the logistics of industrial products, for the purposes of identification and to check delivery processes [30].

*5) Artificial Intelligence (AI):* AI has been employed in smart systems over a long period of time[31], being the science of creating intelligent machines to facilitate everyday life [32]. AI covers many areas, including computer vision, data mining, deep learning, image processing and neural networks [16, 33]. AI technologies are now emerging to assist and improve efficiency and tackle many of the challenges facing the agricultural industry, including soil health, crop yield and herbicide-resistance. According to Sennaar [34], agricultural AI Cloud applications fall into three main categories, as discussed below.

*a) Robots:* these are developed and programmed to handle fundamental agricultural tasks (i.e. harvesting crops) more rapidly and with a higher capacity than human workers. Examples of robotic applications include: (a) See and Spray (i.e. a weed control robot) and (b) Harvest CROO (i.e. a crop harvesting robot). Agricultural robots have the potential to become valuable AI applications, i.e. milking robots.

*b) Monitoring Crop and Soil:* this employs computer vision and deep-learning algorithms for processing captured data by sensors monitoring crop and soil health, i.e. the PEAT machine for diagnosing pests and soil defects, based on deep learning application known as Plantix that identifies potential defects and nutrient deficiencies in the soil. A further example is Trace Genomics, a machine learning based service for diagnosing soil defects and providing soil analysis services to farmers. This uses machine learning to provide farmers with a sense of both the strengths and weaknesses of their soil, with the emphasis being on the prevention of poor crops and optimizing the potential for healthy crop production. A SkySquirrel technology is an example of the use of drones and computer vision for crop analysis.

*c) Predictive Analytics:* This analysis captured data, based on machine learning models capable of tracking and predicting various environmental impacts on crop harvest, i.e. changes in weather. Examples of such AI technologies include (a) aWhere (i.e. prediction of weather and crop sustainability) and (b) Farmshots (i.e. monitoring of crop health and sustainability). Crop and soil monitoring technologies are important applications for addressing issues related to climate change. IoT/AI technologies (such as drone and satellite) that generate a large amount of data on a daily basis have the potential to enable agricultural production to forecast changes and detect opportunities. It is predicted that, over the coming years, IoT and AI applications will attract a considerable degree of interest from large industrial agricultural enterprises [34].

The benefits and advantages of the agricultural use of IoT are as follows: (a) efficiency of input; (b) cost reduction; (c) profitability; (d) sustainability; (e) food safety and environmental protection [35]. However, Ferrández-Pastor, et al. [36] considered SSA to contain a number of barriers potentially hindering its adoption: (a) initial expectations and advantages remaining unfulfilled; (b) complexity of technology and incompatibility of components; (c) a lack of products; and (d) the high cost involved in the establishment and maintenance of such facilities. To ensure the adoption and improvement of smart technologies in the agriculture sector, it is vital for farmers to be trained and given up-to-date knowledge of IoT/AI technologies. Furthermore, it is crucial to test and validate IoT/AI applications, due to the high risk involved in the adoption of these technologies in a critical sector, along with the influence of environmental factors.

*C. Examples of IoT/AI Technologies in Current Agriculture Practices*

There are many types of IoT and AI sensors and applications in current agricultural studies and development. Table 1 provides an overview of the most commonly employed IoT/AI platforms/technologies found in smart agriculture.

*D. Examples of an Existing AI/IoT Research in Smart Agriculture*

There are a number of specific challenges that need to be considered before investing in smart agriculture, primarily those falling into the following categories: (1) hardware; (2) data analysis; (3) maintenance; (4) mobility; and (5) infrastructure [56]. Nonetheless, there are many research efforts in the field of IoT/AI to support the creation and establishment of SSA, as shown in Table 2.

*(IJACSA) International Journal of Advanced Computer Science and Applications,*
*Vol. 10, No. 5, 2019*
TABLE I. EXAMPLE OF IoT/AI APPLICATIONS IN SMART AGRICULTURE

| Category | Tool/Company | Description |
|---|---|---|
| Climate conditions Monitoring | allMETEO [37] | A portal to manage IoT micro weather stations, to gather real-time data access and create a weather map. It also provides an API for easy real-time data transfer into developed or existing infrastructure. |
| | Smart Elements [38] | A collection of products that improve efficiency by eliminating manual checking. They work by deploying a wide range of sensors generating a report back to an online dashboard, allowing rapid and informed decisions based on real-time conditions. |
| | Pycno [39] | A software and sensor allowing continuous data collection and flow from the farm to smartphone. It also contains a dashboard to apply the latest phenological and disease models to monitor trends and assess risk to agricultural products. |
| Greenhouse automation | Farmapp [40] | A process of monitoring pests and diseases, generating reports for mobile applications. It records the data quickly and more efficiently than traditional methods (i.e. paper), allowing a smooth implementation. The stored data is synchronized with the server, enabling the following metrics to be immediately observed: (1) a satellite map with recorded points; (2) the current sanitary status of the farm; (3) comparative heatmaps to easily compare previous measures with the current situation; and (4) charts and reports concerning pests and diseases. |
| | Growlink [41] | A platform that tightly integrates hardware and software products, enabling smarter working, including providing wireless automation and control, data collection, optimization, and monitoring and visualization. |
| | GreenIQ [42] | A system to control irrigation and lighting from all locations and to connect IoT devices to automation platforms. |
| Crop management | Arable [43] | A device that combines weather and plant measurements, sending data to the Cloud for instant retrieval from all locations. It offers continuous indicators of stress, pests and disease. |
| | Semios [44] | A platform focused on yield improvement. It enables farmers to assess and respond to insects, disease and the health of crops using real-time data, forming on-site sensing, big data and predictive analytics solutions for sustained agricultural products. |
| livestock monitoring and management | SCR/Allflex [45] | An advanced animal monitoring system, aimed at the collection and analysis of critical data, including for individual animals. It delivers, when needed, the heat, health and nutrition insights required by farmers for effective decision making. |
| | Cowlar [46] | A smart neck collar for monitoring dairy animals to gather information on temperature, rumination, activity and other behavior. The intelligence algorithm in the system allows for the detection of health disorders before the appearance of visual symptoms. It can monitor body movement patterns and gait to provide accurate oestrus detection alerts. It uses a solar power base unit, along with a waterproof and non-invasive monitoring system, both comfortable for the animal and requiring minimum maintenance. |
| End-to-end farm management systems | FarmLogs [47] | This system monitors field conditions, facilitating the planning and managing of crop production. It also markets agricultural products. |
| | Cropio [48] | A decision-making tool used to optimize fertilization and irrigation to control the amount of fertilizer and reduce the use of water. It combines weather information and satellite data to monitor crops and field forecasts. |
| Predictive Analytics | Farmshots [49] | A system analyzing satellite and drone images of farms fields to map potential sign of diseases, pests and poor nutrition. It turns images into a prescription map to optimize farm production and view analytics on farm performance. Generated data in the Cloud can be exportable into nearly all agricultural software for prescription creation. |
| | aWhere [50] | A platform employed for weather prediction and information on crop sustainability. Its goal is to deliver complete information and insight for real-time agricultural decisions on a daily basis and at a global level. |
| Crop and Soil Health Monitoring | Plantix [51] | A machine learning based tool to control and manage the agriculture process, disease control, and the cultivation of high-quality crops. |
| | Trace Genomics [52] | A soil monitoring system performing complex tests (i.e. DNA) on soil samples. It uses a machine learning process known as 'genome sequencing' that generates a health report for a soil sample by reading its DNA and comparing it to a large soil DNA database. |
| Agriculture machines/drones | SkySquirrel [53] | A drone system aimed at helping users to improve their crop yield and reduce costs. Users pre-program a drone's route, and, once deployed, the device will leverage computer vision to record images to be used for analysis. Once the drone completes its route, users can transfer the data to a computer and upload it to a Cloud drive. It uses algorithms to integrate and analyze the captured images and data to provide a detailed report on the health and condition of crops. |
| | See & Spray [54] | A robot designed to control weeds and protect crops. It leverages computer vision to monitor and precisely spray weeds and infected plants. |
| | CROO [55] | A robot that assists in the picking and packing of crops. The manufacturer claimed that this robot can harvest eight acres in a single day and replace the work of thirty human laborers. |

96 | P a g e
www.ijacsa.thesai.org

TABLE II. IoT/AI Research and Development in Smart Sustainable Agriculture

| Researcher/s | Year | Summary |
|---|---|---|
| Ray [57] | 2017 | The researchers undertook a review of various potential IoT applications, including the specific issues and challenges associated with IoT deployment to improve farming. They comprehensively analyzed the specific requirements the devices and wireless communication technologies associated with agricultural IoT applications. They presented different case studies to explore existing IoT based solutions operated by various organizations and individuals, followed by categorizing them based on their deployment parameters. Furthermore, they identified a number of factors for the improvement and future road map of work using IoT. |
| Mekala and Viswanathan [58] | 2017 | The researchers surveyed a number of conventional applications of Agricultural IoT Sensor Monitoring Network technologies utilizing CC. Their study aimed at understanding the diverse technologies to build smart, sustainable agriculture. They addressed a simple IoT agriculture model with a wireless network. |
| Kamilaris, et al. [59] | 2017 | The researchers reviewed work in agriculture employing the practice of big data analysis to solve various different problems. Their review emphasized the opportunities provided by big data analysis for the development of smarter agriculture, the availability of hardware and software, as well as the techniques and methods for big data analysis. |
| Rajeswari, et al. [29] | 2017 | The researchers investigated a number of different features, i.e. humidity, temperature sensing, server-based remote monitoring system detection and soil moisture sensing. They used sensor networks to measure temperature, moisture and humidity in place of manual checking. They deployed several sensors in different locations within farms, using a single controller. Their major objective was to collect real-time data of the agriculture production environment to establish an easy access agricultural advice, in order to identify weather or crops patterns. |
| Antonacci, et al. [5] | 2018 | The researchers attempted to provide nanotechnology-based (bio) sensors to support farmers in delivering an analysis that is accurate, fast, cost-effective, and useful in the field to identify water and soil nutrients/pesticides, soil humidity, and plant pathogens. |
| Cadavid, et al. [60] | 2018 | The researchers proposed an extension to a popular open-source IoT platform, known as 'Thingsboard'. This formed the core of a Cloud-based Smart Farming platform and deliberate sensors, a decision support system, and a configuration of remotely autonomous and controlled machines (e.g. water dispensers, rovers or drones). |
| Soto-Romero, et al. [61] | 2019 | The researchers designed an easily insertable cylindrical sensor with internal electronics to offer a low power electronic architecture to measure and communicate wirelessly with a LoRa, Sigfox network or mobile phones. |
| Nóbrega, et al. [62] | 2019 | The researchers reviewed the proposed stack and details of the recent developments within smart agriculture, focusing on IoT/Machine-2-Machine interaction. They described the design and deployment of a gateway addressing the requirements of the SheepIT service, evaluating this gateway using real scenarios in terms of performance, thus demonstrating its feasibility and scalability. |

## III. METHODOLOGY

As well as undertaking the literature review, the current researcher enhanced this study by informally interviewing experienced farmers. The study aims to establish an IoT/AI SSA architecture, as well as exploring the potential of the use of IoT and AI as a backbone to establish an SSA platform. A review was employed to identify, analyze and study key books, journals, reports, and white papers, in order to achieve the above-noted aim. The lack of existing studies in this area ensures that this current research also contributes to the body of knowledge by establishing an IoT/AI framework for the adoption of smart technologies, in order to establish smart sustainable agricultural practices. Fig. 1 shows the methodology adopted for this study.

## IV. RESULTS AND DISCUSSION

Based on the research aim outlined in the Methodology, this section is divided into (A) Domains of Smart Sustainable agricultural model; (B) B. Proposed IOT/AI SSA platform as a solution; and (C) Proposed IoT/AI technical architecture for SSA platform.

### A. Domains of Smart Sustainable Agricultural Model

The results from the literature review revealed that several domains need to be considered when adopting the smart agricultural model. Fig. 2 demonstrates the interrelation and complexity of data flow between different Smart, Sustainable Agriculture domains.

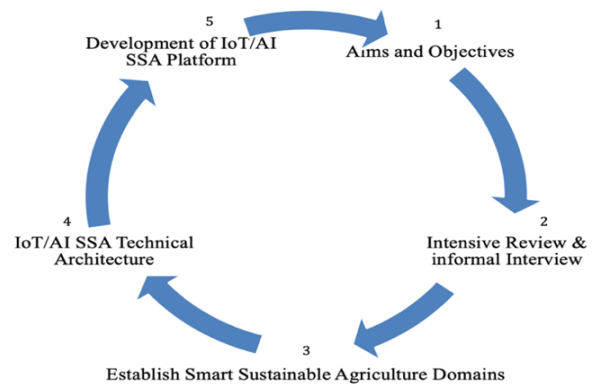

Fig. 1. IoT/SSA Research Methodology.

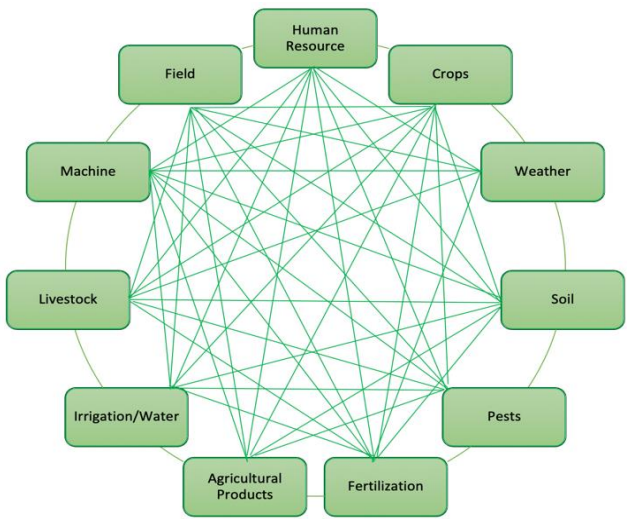

Fig. 2. Chaotic Data Flow and Interrelation of SSA Domains.

These domains are discussed individually, as follows:

- *Human resources:* This refers to people, policies and practices within the agricultural environment, which are as important as in any other domain, as are weather and technology. Human resources receive careful attention, due to their significant impact on production, as well as financial and marketing decisions. Whatever its size, an agricultural concern requires effective human resources management and planning, including hiring and keeping employees who are engaged, high-performing and effective communicators. Providing up-to-date knowledge potentially opens the means to adopt smart technologies in an agricultural environment.

- *Crops*: This refers to a plant that can be grown and harvested extensively for subsistence or profit: (1) food crops (i.e. for human consumption); (2) feed crops (i.e. for livestock consumption); (3) fibre crops (i.e. for cordage and textiles); and (4) oil crops (i.e. for consumption or industrial uses).

- *Weather:* This plays a major role in determining the success of agricultural processes. Most field crops and livestock are solely dependent on climatic conditions to provide life-sustaining water and energy. Adverse weather can cause losses in agricultural products, particularly during critical stages of growth. The elements of weather (solar radiation, temperature, precipitation, humidity and wind) influence the physiology and production of agricultural plants and animals. Severe weather (i.e. tornadoes, drought, flooding, hail and wind storms) can cause considerable damage and destruction to fields and livestock.

- *Soil:* This forms a critical aspect of successful agriculture, being the source of nutrients used to grow crops, which are subsequently passed into plants and then to humans and animals. Healthy soils produce healthy and rich food supplies; however the health of soil tends to decline over time, forcing farmers to move to new fields. Soil health depends on regional conditions and climates, with soil nutrients more likely to deteriorate in dry climates, particularly if irrigated, which, if not managed carefully, can result in salinization, i.e. a build-up in the level of salts and chemicals contained in water. Healthy and rich soil can be achieved through the use of IoT sensors to monitor its chemical status, using specific sensors (e.g. moisture sensors), whose data readings are transferred to the data management and analysis layer for analysis, assisting decisions concerning the need for fertilizer.

- *Pests:* These consist of any living creature that is invasive, or damaging to crops, livestock or human structures. Pests often occur in high quantities, to the detriment of agricultural products. It is vital to control and monitor these creatures by means of IoT/AI technologies, to avoid serious diseases, including plague and malaria, as well as plant and livestock diseases.

- *Fertilization:* Soils naturally contain many nutrients, i.e. nitrogen, phosphorous, calcium and potassium. Crops are unable to function effectively and produce high quality food when their nutrient level is low. The natural levels of nutrients in the soil need to be enhanced by the addition of nutrients once crops have been harvested. Fertilizers have been used since the beginning of agriculture, but it is now recognized that their extensive use can, if not correctly controlled, harm the environment. Therefore, farmers use IoT sensors to read and test soil samples for baseline testing to enable them to add fertilizers using correct and appropriate measurements. Fertilization is an important method of maintaining sustainable agricultural production systems.

- *Agricultural Products:* These are derived from cultivating crops or livestock to sustain or enhance human life. Human beings also use a wide collection of agricultural products on a daily basis, i.e. food and clothing. Agricultural products fall into the following groups: (a) grains; (b) foods; (c) fuels; (d) fibres; (e) livestock; and (f) raw materials. Food is the most extensively produced agricultural product.

- *Irrigation/Water:* Water demand in agriculture is now rising globally and particularly in Mediterranean countries, increasing the pressure to preserve available freshwater resources. Smart, sustainable agriculture processes should therefore focus on new and efficient techniques to improve agricultural productivity, which promote considerable savings in terms of food consumption and wasted water.

- *Livestock:* These are animals raised in a domesticated agricultural environment, for the purposes of labour and to produce commodities such as eggs, meat, milk, fur, wool and leather. Animal husbandry is a component of current agriculture and refers to the breeding, maintenance and slaughter of livestock.

- *Machines:* Agricultural equipment is any kind of machinery used to assist with farming. Such machines

can be light or heavy, i.e. tractors. Modern farm machinery is seen as the important driver for increased agricultural sustainability, efficiency and competitiveness. Smart technologies can reduce the impact of farming practices within global agriculture. The current development of agricultural machinery addresses environmental challenges, while increasing productivity and bringing economic benefits. These smart agricultural machines should be: (a) fast, accurate, versatile and intelligent; (b) produce less $CO^2$ emissions and (c) make use of bioenergy.

- *Fields:* This refers to an area of land used for agricultural purposes, i.e. crops, cultivation or for livestock. Many fields have borders composed of a strip of bushes used to provide both food and cover, in order to ensure the survival of wildlife. Monitoring field activities using IoT devices can have a significantly positive impact on controlling objects within the field.

B. *Proposed IOT/AI SSA Platform as a Solution*

This platform would prove a valuable medium to facilitate data flow and sharing among SSA domains. Many researchers have developed IoT architectures, but their efforts have tended to target specific areas of IoT/AI, i.e. sensors or weather monitoring systems [58, 60, 61]. This current paper is proposing a holistic IoT/AI platform to cover all areas within a SSA environment, performing the following tasks: (a) manage and govern data flow between SSA domains; (b) facilitate the integration of the different components of SSA architecture; (c) tackle interoperability issues caused by the utilization of different tools and software; (d) provide easy-to-use interfaces for interaction; (e) provide an ability to generate reports based on real-time data and keep it updated; (f) store generated data in sustainable storage place (i.e. the Cloud), to enable it to permanently recorded for future reuse; (g) isolate different layers to improve the development process in the future; and (h) the platform should consist of several nodes, so that, in the case of any failure, other nodes can keep the system up and running. Fig. 3 shows how the SSA-IoT/AI platform would be used at the center of the SSA domain to facilitate business process and data flow and to share within a smart, sustainable agricultural environment.

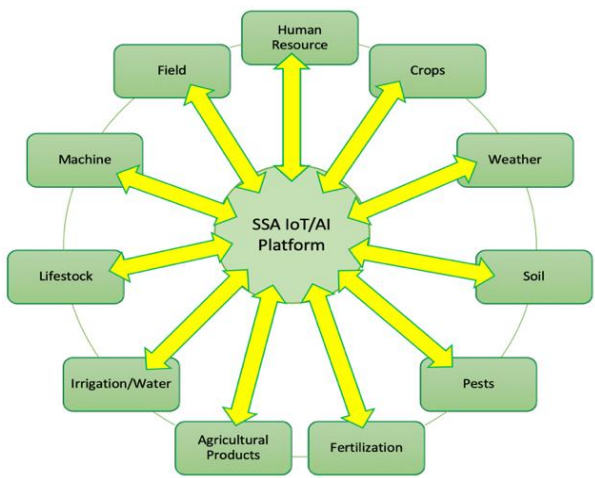

Fig. 3. Proposed SSA-IoT/AI Platform.

C. *Proposed IoT/AI Technical Architecture for SSA Platform*

Fig. 4 shows the overall AI/IoT technical architecture for SSA. It consists of two main components: (a) the first component: SSA Layers and AI/IoT technologies; and (b) the second component: data lifecycle within SSA architecture and data process location. In order to provide additional detail concerning the framework, the following description gives further details about each component:

*1) The First component: Smart Sustainable Agriculture (SSA) layers and (SSA) AI/IoT technologies*

The first component of smart, sustainable agriculture (SSA), AI/IOT framework, consists of the following layers: (a) Physical Hardware and Storage layer; (b) AI and Data Management layer, and Governance layer; (c) Network layer; (d) Security layer; (e) Application layer; (f) IoT and sensing layer; and (g) SSA domain layer. Fig. 4 demonstrates the following description of each layer, highlighting its primary role:

- *Physical hardware and Storage layer*: This layer consists of powerful hardware to host virtualized machines, as well as dedicated traditional storage medium, cloud storage solutions or hybrid storage solutions. It contains the supporting hardware for IoT devices that exist in Internet of Thins and Sensing layer.

- *Artificial Intelligence and Data Management Layer*: This layer is responsible for managing processes and controlling the business logic, focussing on three main activities: (i) data analysis and processing, using data mining and intelligence statistical analysis on generated data; (ii) data classification and transformation, using ontologies, and machine learning to classify and transform analysed data; and (iii) data interpretation, representing the transformed data into knowledge to make machines smarter.

- *Network Layer:* This layer contains all the different types of network connections exist, with this cloud consisting of wired or wireless network connections/devices. It is important that this layer uses the latest networking technologies to keep up with the most recent developments in other sectors. Its main job is to facilitate the transaction of all data to and from different layers within the architecture. The technologies of IoT in this layer include Internet, WIFI and GSM/CMDA. It is responsible for data accessibility and availability throughout other layers. Further, it manages data and its flow within between all layers.

- *Security Layer:* This layer is responsible for data security transferred among different layers and should be the means of addressing any security concerns and vulnerabilities within all other IoT/SSA layers, i.e. malware, spyware and viruses. It should employ up-to-date security solutions and AI algorithms to block and quarantine any threat to the platform.

- *Applications Layers:* This layer gathers different applications related to smart sustainable agriculture. It is built based on AI and data management layers. Many smart, sustainable agricultural applications could be developed and integrated into this layer, i.e. crop monitoring applications and drone-controlled applications. This layer focuses on the supervision aspect of the data flow and migration between all layers and can provide an authorized institution with full or partial governance on data migration, transactions and access.

- *Internet of Things (IoT) and Sensing Layer:* This forms the first interaction layer with SSA domains. It uses and hosts various types of IoT devices (e.g. sensors), capable of collecting data from real-world objects, sharing it to provide real-time data. Many sensors in the Cloud are hosted and integrated within this layer, i.e. humidity sensors, moisture sensors and weather monitoring systems. Furthermore, this layer is responsible for operating robotic and drone actuators to assist in the mobility of intelligent machines within the agricultural area. It thus allows intelligent machines to move between locations, in order to cover a wide area.

- *SSA Domain Layer:* This layer hosts various different Smart Sustainable Architectural domains and forms the main source for data generated from various agricultural domains, including: fields; machines; human resources; and crops. It forms the basis for IOT/AI SSA platforms, as it contains various raw data formats without interference.

*2) The Second component: data lifecycle within SSA Architecture and its process location:*

Fig. 5 demonstrates that data lifecycle remains in line with SSA Architecture layers. This commences with the original data source, i.e. SSA domain layers. The acquisition and capturing of data is undertaken at the layer containing sensors and actuators. These captured data are then passed to the application layer for business logic and control, following which, the data must be checked for security issues before moving to data analysis and processing. This is followed by data classification and transformation of the analyzed and processed data before it moves to data interpretation and the resulting building decisions. The final stage is to store the data for future retrieval.

There are two locations for sharing and processing of generated data. Firstly, on-site. Here, the generated data is more likely to be shared and processed within the location of the agricultural field, covering the four architecture layers: (1) SSA; (2) the domain layer; (3) the IoT layer; and (4) the application layer. Secondly, off-site: here the sharing and processing of the generated data must be outsourced to the physical location 'data centre', in which data is processed and analyzed away from the field. It can cover the four architecture layers: (1) network; (2) AI and data management; (3) physical; and (4) storage. It is also important to highlight that the Security layer forms a common layer between on-site and off-site processing locations.

Fig. 4. Overall AI/IoT Platform Technical Architecture for SSA.

Fig. 5. IoT/AI Data Lifecycle.

V. Conclusion

This paper has established the importance of employing recent and advanced computing technologies in the agricultural sector, in particularly AI and IoT. Agriculture is considered central to the survival of human beings. Supporting the current practices of traditional agriculture with recent IoT/AI technologies can improve the performance, quality and volume of production. This study has reviewed the existing IoT/AI technologies discussed within the main research journals in the area of agricultural. Furthermore, it categorized the main domains of smart, sustainable agriculture, i.e. human resources; crops; weather; soil; pests; fertilization; farming products; irrigation/water; livestock; machines; and fields. The major contribution of this paper concerns the AI/IoT technical architecture for SSA, leading to an emphasis on the research and development of a unified AI/IoT platform for SSA, to positively resolve issues resulting from the fragmentary nature of the agricultural process. Future work will include investigation of the process of implementing AI/IoT technologies for SSA by applying the proposed AI/IoT technical architecture in the form of the prototype of a unified platform on real test cases. This will identify the relevant strengthens and weaknesses for further improvement and enhancement.